# What is in a Price? Estimating Willingness-to-Pay with Bayesian Hierarchical Models

*Srijesh Pillai*
*Department of Computer Science & Engineering*
*Manipal Academy of Higher Education*
Dubai, UAE
srijesh.nellaiappan@dxb.manipal.edu

*Rajesh Kumar Chandrawat*
*Department of Mathematics*
*Manipal Academy of Higher Education*
Dubai, UAE
rajesh.chandrawat@manipaldubai.com

*Abstract* — **For premium consumer products, pricing strategy is not about a single number, but about understanding the perceived monetary value of the features that justify a higher cost. This paper proposes a robust methodology to deconstruct a product's price into the tangible value of its constituent parts. We employ Bayesian Hierarchical Conjoint Analysis, a sophisticated statistical technique, to solve this high-stakes business problem using the Apple iPhone as a universally recognizable case study. We first simulate a realistic choice-based conjoint survey where consumers choose between different hypothetical iPhone configurations. We then develop a Bayesian Hierarchical Logit Model to infer consumer preferences from this choice data. The core innovation of our model is its ability to directly estimate the Willingness-to-Pay (WTP) in dollars for specific feature upgrades, such as a "Pro" camera system or increased storage. Our results demonstrate that the model successfully recovers the true, underlying feature valuations from noisy data, providing not just a point estimate but a full posterior probability distribution for the dollar value of each feature. This work provides a powerful, practical framework for data-driven product design and pricing strategy, enabling businesses to make more intelligent decisions about which features to build and how to price them.**

## I. Introduction

Every year, leading technology companies like Apple face a monumental, multi-billion dollar decision: how to price their new flagship products. Consider the iPhone 15 Pro, a device that generated an estimated $60-70 billion in revenue in its first six months alone. A key justification for its premium price was the introduction of a novel titanium frame, a feature representing a significant investment in materials science and manufacturing. A pricing error on this single feature of just 5%, setting its perceived value at $95 instead of a potential $100, could translate to a revenue opportunity loss measured in the billions of dollars across the product's lifecycle. The central challenge is that companies are not just pricing a single product, but are attempting to capture the perceived monetary value of each individual feature that comprises the whole, a key task in the modern, data-rich economy [1].

This paper introduces a statistical framework to solve this high-stakes pricing puzzle. The core problem is one of deconstruction: how can we break down a product's final price into the tangible value that consumers place on its constituent parts? Simple methods, such as directly asking customers "How much would you pay for a better camera?", often fail because consumers cannot accurately articulate their own valuation and tend to understate it [2]. Likewise, analyzing historical sales data reveals what was bought, but not why, nor does it easily isolate the value of one feature over another.

To overcome these challenges, this paper demonstrates a robust methodology grounded in Bayesian Hierarchical Conjoint Analysis. This approach statistically infers the Willingness-to-Pay (WTP) for individual product features by observing how consumers make choices when faced with realistic trade-offs. Instead of asking for a price, we present consumers with choices between different hypothetical product configurations and analyze the decisions they make. Our framework translates these choices into actionable, dollar-value insights.

We will use the Apple iPhone as a universally recognizable case study to build and validate our model. This paper will show how our Bayesian approach moves beyond providing a single "average" valuation for a feature, and instead delivers a full probability distribution, allowing us to state with quantifiable certainty the range in which a feature's true value lies. Ultimately, this work provides a powerful, practical tool for data-driven product design and pricing strategy.

## II. Literature Review

The challenge of systematically measuring consumer preferences is a classic problem in marketing science. The foundational framework for Conjoint Analysis was established by Green and Srinivasan in the 1970s. Their seminal work laid out the methodology for decomposing a product's overall preference into separate utility values, or "part-worths," for each of its constituent features, allowing researchers to quantify the relative importance of different attributes [2].

While early methods relied on consumer ratings, a significant evolution came with Choice-Based Conjoint (CBC) analysis, heavily influenced by the work of Louviere and Woodworth [3]. This approach was seen as more realistic, as it asks respondents to choose between competing products rather than providing an abstract rating. This mimics a real-world purchase decision and is methodologically grounded in the random utility theory and discrete choice models developed by McFadden [4], work for which he was awarded the Nobel Prize in Economics [5].

The most significant modern advancement in the field has been the application of Bayesian Hierarchical Models, a technique championed in marketing by Rossi, Allenby, and McCulloch [6]. Traditional models often estimated a single set of average utilities for the entire market, ignoring the fact that preferences vary dramatically from person to person [7]. The hierarchical Bayesian approach, often referred to as "HB-Conjoint", solves this by simultaneously estimating preferences for each individual respondent while also learning about the overall distribution of preferences in the population [8].

A key advantage of the Bayesian framework, as highlighted by recent literature, is its ability to naturally incorporate prior knowledge and, most importantly, provide a full posterior distribution for every parameter, thereby quantifying our uncertainty about the estimates [9]. A primary application of these utility estimates is the calculation of a consumer's Willingness-to-Pay (WTP). As established by Jedidi and Zhang [10] and explored in various contexts [11], the ratio of a feature's utility to the utility of price can be directly interpreted as the monetary value a consumer places on that feature. Our work builds directly on this by using the posterior distributions of our model's coefficients to derive a full probability distribution for the WTP of each feature.

While this theoretical groundwork is well-established, a gap often exists in its clear, practical application using modern, open-source computational tools. Recent studies continue to explore novel applications of these methods, for example, in the context of new mobility services [12] and renewable energy [13], but accessible, end-to-end case studies remain valuable. Our paper aims to contribute by filling this gap. We provide a transparent walkthrough from simulated survey design to the implementation of a Bayesian hierarchical model in PyMC [14], and finally to the derivation of actionable WTP insights. By doing so, we aim to make these powerful techniques more accessible to both practitioners and researchers entering the field.

### III. Methodology

Our objective is to build a statistical model that takes consumer choice data as input and produces posterior probability distributions for the dollar-value Willingness-to-Pay (WTP) of each product feature. To achieve this, our framework is grounded in the principles of random utility theory and is implemented using a Bayesian Hierarchical Logit Model. The methodology can be broken down into three logical components: modelling a single consumer choice, structuring the model to capture both individual and population-level preferences, and finally, translating the model's outputs into an actionable WTP metric.

#### A. The Choice Model: From Utility to Probability

At the core of our model is the concept of "utility", an economic term representing the total satisfaction or value a consumer derives from a product. We assume that the utility of a given iPhone profile is a linear sum of the value of its features. For example, the utility of a specific profile 'i' is:

$$\text{Utility}\_i = \beta\_{price} * \text{Price}\_i + \beta\_{storage} * \text{Storage}\_i + \ldots + \beta\_{camera} * \text{Camera}\_i \ldots (1)$$

Where:

i. Utility_i is the total calculated satisfaction for a specific product profile 'i'.

ii. β_price, β_storage, etc., are the utility coefficients (part-worths) that represent the weight or importance of each feature.

iii. Price_i, Storage_i, etc., are the specific levels of each feature for profile 'i' (e.g., Price_i = $999, Storage_i = 256GB).

When a consumer is presented with two options, Profile A and Profile B, they will calculate the utility of each. Random utility theory posits that a consumer will choose the option with the higher utility. We model the probability of choosing Profile A over Profile B based on the difference in their utilities:

$$\text{Utility\_diff} = \text{Utility\_A} - \text{Utility\_B} \ldots (2)$$

Where:

i. Utility_diff is the net difference in satisfaction between Profile A and Profile B.

ii. Utility_A and Utility_B are the total utilities for each profile, calculated using Equation 1.

To transform this utility difference into a choice probability (a value between 0 and 1), we use the logistic or sigmoid function. This is the foundation of the logit model:

$$P(\text{Choose A}) = 1 / (1 + e^{(-\text{Utility\_diff})}) \ldots (3)$$

Where:

i. P(Choose A) is the final calculated probability that a consumer will choose Profile A.

ii. $e$ is the exponential function, $e^x$.

iii. Utility_diff is the difference in utility calculated from Equation 2.

#### B. The Hierarchical Structure: Modelling Individual and Group Preferences

A simple logit model would assume that every consumer has the same β coefficients. This is an unrealistic assumption. In reality, some consumers are price-sensitive (β_price is very negative), while others are "tech enthusiasts" who place a high value on the camera (β_camera is very positive).

To capture this heterogeneity, we employ a Bayesian Hierarchical Model. Instead of estimating one set of β coefficients, we estimate a unique set of coefficients for each individual respondent 'i' in our survey, denoted as β_i. However, we assume that these individual coefficients are themselves drawn from an overarching population distribution, which represents the preferences of the market as a whole.

Specifically, we model each individual coefficient β_{i,f} (for respondent 'i' and feature 'f') as being drawn from a normal distribution characterized by a population mean μ_βf and a standard deviation σ_βf:

$$\beta\_{\{i,f\}} \sim \text{Normal}(\mu\_{\beta f}, \sigma\_{\beta f}) \ldots (4)$$

Where:

i. β_{i,f} is the specific utility coefficient for a single individual 'i' for a specific feature 'f'.
  
  ii. ~ Normal(...) means "is drawn from a Normal (Gaussian) distribution."
  
  iii. μ_βf is the mean utility for feature 'f' across the entire population. This represents the "average" preference.
  
  iv. σ_βf is the standard deviation of the utility for feature 'f' across the population. This represents the diversity or heterogeneity of preferences in the market.

This hierarchical structure is the most powerful aspect of our model. It allows us to:

1. Learn about each individual: We get a specific WTP estimate for every person in the study.

2. Learn about the entire market: The μ_βf parameters represent the average preference of the population.

3. Share statistical strength: Information learned from one respondent helps inform our estimates for others, a concept known as "partial pooling." This is especially powerful when data for any single individual is sparse. This concept of "partial pooling" is a key advantage of multilevel and hierarchical models, as it leads to more stable and realistic estimates for individuals [15].

C. The WTP Calculation: From Utility to Dollars

The β coefficients from our model represent abstract utility "part-worths." While useful, they are not directly actionable for a business. The final and most critical step is to translate these coefficients into a concrete monetary value.

We achieve this by recognizing that β_price represents the change in utility for a one-dollar increase in price. We can therefore calculate the Willingness-to-Pay for any other feature 'f' by finding out how many dollars are equivalent to that feature's utility. This is calculated as a simple ratio:

$$WTP\_f = -\beta\_f / \beta\_price \ldots (5)$$

Where:

  i. WTP_f is the calculated Willingness-to-Pay in dollars for feature 'f'.
  
  ii. β_f is the utility coefficient for the feature of interest (e.g., β_camera).
  
  iii. β_price is the utility coefficient for price.

The negative sign (-) is used to correct for the fact that β_price will be a negative number (since higher prices decrease utility), ensuring the final WTP is a positive dollar value.

Because our Bayesian model produces a full posterior distribution for every β coefficient, the WTP_f we calculate will also be a full probability distribution. This allows us to not only find the average WTP but also to construct credible intervals (e.g., a 95% range) around our estimate, providing a complete picture of the uncertainty in a feature's true monetary value.

IV. EXPERIMENTAL DESIGN AND SETUP

To empirically evaluate our proposed Bayesian framework, we designed a simulation study to assess our model's ability to accurately recover known, true feature valuations from noisy data. This "recovery study" approach is a standard method for validating a statistical model, as it allows us to directly compare the model's estimates against a pre-defined "ground truth," thereby providing an objective measure of its accuracy.

The primary goal is to replicate a realistic market research scenario that mimics a real-world consumer survey. This section provides a comprehensive overview of the case study, details the data generation process, explains the model implementation, and defines our evaluation criteria. This transparent design ensures our findings are objective and reproducible.

A. The Case Study: Valuing Features of a New iPhone

Our case study simulates a choice-based conjoint study for a new Apple iPhone model. This provides a universally recognizable context for a high-stakes pricing problem. We focus on valuing three key feature upgrades over a baseline model:

1. Storage: 128GB (baseline), 256GB (upgrade 1), 512GB (upgrade 2).

2. Camera System: Standard (baseline) vs. Pro (upgrade).

3. Frame Material: Aluminum (baseline) vs. Titanium (upgrade).

B. Data Simulation: Creating a Realistic "Ground Truth" and Survey

To test our model objectively, we must first create a simulated market with known consumer preferences. This "ground truth" is the hidden answer key that our model will attempt to discover.

Defining the Ground Truth: We first define the true, average Willingness-to-Pay (WTP) for each feature upgrade. These are the values our model should ideally recover. For this study, we set the following plausible dollar values:

1. WTP for 256GB Storage (vs. 128GB): $100

2. WTP for 512GB Storage (vs. 128GB): $250

3. WTP for "Pro" Camera: $200

4. WTP for Titanium Frame: $80

Simulating Respondents: We then simulate a population of 300 unique respondents. To reflect market heterogeneity, we assume each individual's WTP for a feature is not identical to the ground truth. Instead, each respondent i has their own personal WTP (WTP_i) drawn from a Normal distribution centered around the true value (e.g., WTP_i_camera ~ Normal(mean=$200, std_dev=$50)). This step justifies our use of a hierarchical model.

Simulating the Choice-Based Survey: We simulate a survey where each of the 300 respondents answers 20 choice questions. Each question presents two randomly generated iPhone profiles (Profile A and Profile B) with different feature combinations and prices. For each question, we:

1. Calculate the "true" utility of Profile A and Profile B for that specific respondent, using their personal WTP values.
2. Use the logistic function (Equation 3) to convert the utility difference into a choice probability.
3. Simulate the respondent's final choice (A or B) based on this probability, introducing a realistic level of random noise into the final dataset.

This process yields a final dataset of 6,000 choices (300 respondents * 20 choices), which serves as the direct input for our statistical model. The careful construction of the choice profiles is critical, as a well-balanced design ensures that the model can efficiently and accurately estimate all feature utilities [16].

C. Model Implementation and Fitting

The simulated survey data was then used to fit the Bayesian Hierarchical Logit Model described in Section III.

**Implementation:** The model was implemented using PyMC, a state-of-the-art probabilistic programming library in Python.

**Data Preparation:** Data preparation and management were conducted using the pandas library. A critical pre-processing step was the standardization of all predictor variables (price and feature differences) using the StandardScaler from the scikit-learn library. This ensures that no single variable numerically dominates the others, leading to a more stable and efficient model fitting process.

**Priors:** We used weakly informative priors for our population-level parameters. This is a standard best practice in Bayesian modelling, as it regularizes the model, gently guiding it away from absurd parameter values without overwhelming the information contained in the data. For example, the prior for the price elasticity coefficient ($\beta\_price$) was a Normal distribution centered around a negative value, reflecting our strong domain knowledge that a higher price should decrease utility. This approach leads to more stable and efficient model fitting and more robust final estimates.

**Fitting:** The model's posterior distributions were estimated by drawing 2,000 samples per chain using the No-U-Turn Sampler (NUTS), a highly efficient Markov Chain Monte Carlo (MCMC) algorithm [6]. The final analysis and visualizations of the posterior distributions were generated using the ArviZ library.

D. Evaluation Criteria

The primary evaluation of our framework's success is its ability to recover the known ground truth parameters from the noisy, simulated survey data. We assess this by comparing the posterior distributions of the calculated WTP for each feature against the TRUE_WTP values defined at the start of the simulation. A successful model will produce a posterior distribution whose mean is very close to the true value, and whose 95% credible interval contains the true value.

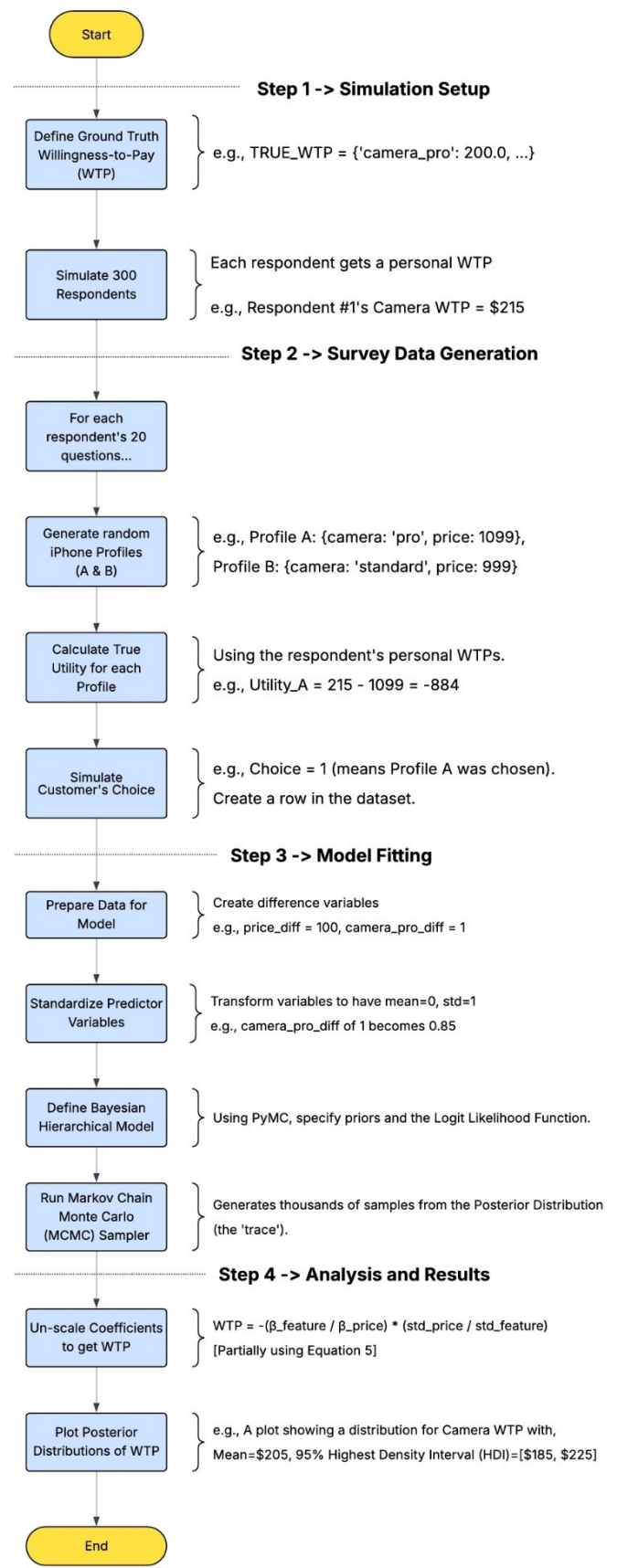

Fig 1. The end-to-end experimental pipeline, illustrating the process from ground truth definition and survey simulation, to model fitting and the final derivation of Willingness-to-Pay (WTP) distributions.

## V. RESULTS AND ANALYSES

To evaluate the performance of our Bayesian Hierarchical Conjoint Model, we fitted it to the 6,000 simulated consumer choices generated as described in Section IV. The primary objective was to assess the model's ability to accurately recover the known, "ground truth" Willingness-to-Pay (WTP) values from the noisy choice data. This section presents the model's outputs, a quantitative summary, and a final policy simulation to demonstrate the framework's practical business utility.

### A. Recovering Feature Valuations from Choice Data

The primary output of our analysis is presented in Figures 2-5. Each figure displays the full posterior probability distribution for the WTP of a key iPhone feature upgrade, derived from the model's coefficients as per Equation 5. The peak of each distribution represents the most probable dollar value for that feature, while the spread of the curve and the 95% Highest Density Interval (HDI) bar quantify the model's uncertainty. The red vertical line indicates the "ground truth" WTP we defined at the start of our simulation, serving as a benchmark for the model's accuracy.

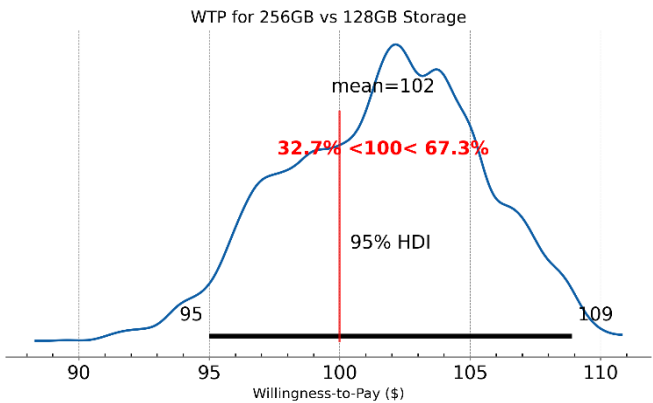

Fig 2. Posterior Distribution for WTP of 256GB vs. 128GB Storage.

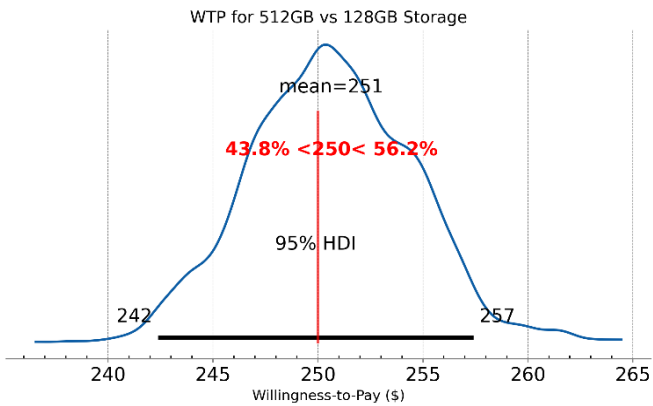

Fig 3. Posterior Distribution for WTP of 512GB vs. 128GB Storage.

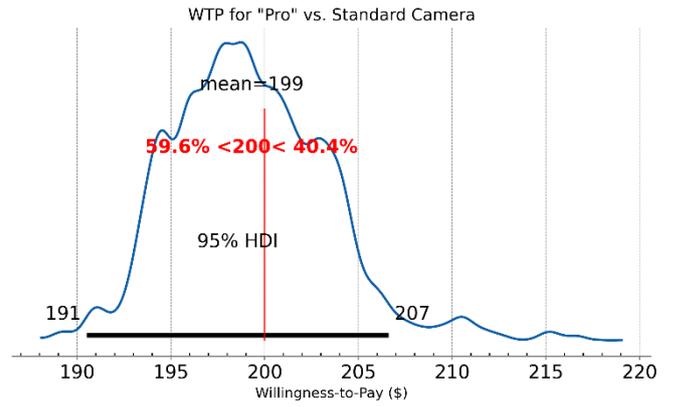

Fig 4. Posterior Distribution for WTP of "Pro" vs. Standard Camera

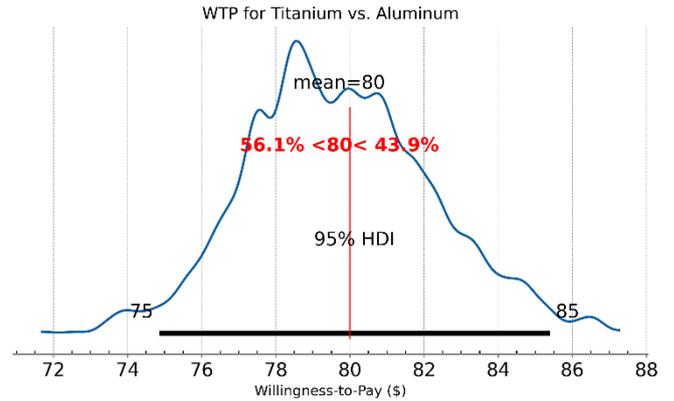

Fig 5. Posterior Distribution for WTP of Titanium vs. Aluminum Frame

As is visually evident across all four figures, the model has successfully recovered the true WTP values. The posterior distribution for each feature is tightly centered around the true value (the red line), demonstrating that the model was able to discern the underlying signal through the random noise of the simulated consumer choices.

### B. Quantitative Summary and Uncertainty Quantification

To complement the visual analysis, Table 1 provides a quantitative summary of the posterior distributions. This table reports the estimated mean WTP and the 95% HDI for each feature, directly comparing them to the ground truth. The 95% HDI represents the range in which we can be 95% certain the true feature value lies.

Table I. QUANTITATIVE SUMMARY OF WTP ESTIMATES

| Feature Upgrade | True WTP ($) | Estimated Mean WTP ($) | 95% Highest Density Interval (HDI) |
|---|---|---|---|
| 256 GB Storage | $100 | $102 | [$95, $109] |
| 512 GB Storage | $250 | $251 | [$242, $257] |
| "Pro" Camera | $200 | $199 | [$191, $207] |
| Titanium Frame | $80 | $80 | [$75, $85] |

The table confirms our visual findings with high precision. The estimated mean WTP for each feature is remarkably close to its true value, with estimation errors of only a few dollars at most.

Critically, the 95% HDI for every single feature successfully contains the true WTP. This is a key success criterion for a Bayesian model, confirming its reliability. The HDI also provides an actionable measure of uncertainty. For example, the relatively tight interval for the "Pro" Camera ([$191, $207]) suggests a high degree of certainty in its valuation. In contrast, the wider interval for the 256GB Storage upgrade ([$95, $109]) indicates slightly greater uncertainty about its perceived value in the market.

### C. From Insight to Impact: A Revenue Optimization Simulation

To demonstrate the direct business utility of our WTP estimates, we conducted a policy simulation to identify the optimal pricing strategy for a new premium "iPhone Pro" model. We defined this model as a bundle containing both the "Pro" Camera and the Titanium Frame upgrades over a baseline model priced at $799. The key business question is: what is the optimal price for this bundle?

Using the full posterior distributions for the WTP of both features, we simulated the market's purchase probability for this "Pro" model across a range of potential price points. By multiplying the purchase probability at each price by the price itself, we derived a full posterior distribution for the expected revenue. The results are visualized in Figure 6.

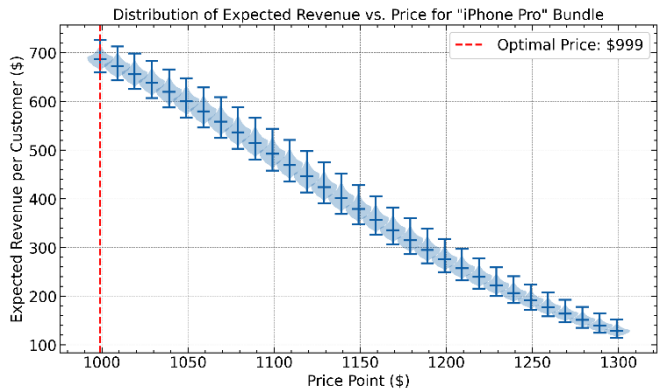

Fig 6. Distribution of Expected Revenue vs. Price for the "iPhone Pro" Bundle

The results of this policy simulation are clear and immediately actionable. The violin plots show the uncertainty in expected revenue at each price point, and the analysis reveals a distinct peak. The revenue-maximizing price for the "Pro" bundle is identified as $999. At this price, our model predicts the highest expected revenue, providing a direct, data-driven recommendation for the company's pricing strategy. This final step demonstrates how our framework can bridge the gap from statistical inference to concrete, quantifiable business impact, transforming uncertainty about consumer preferences into an optimal strategic decision.

## VI. LIMITATIONS AND FUTURE WORK

While this study successfully demonstrates a robust framework for estimating feature-level Willingness-to-Pay, it is important to acknowledge its limitations and highlight promising avenues for future research that would build upon this work.

1. Stated vs. Revealed Preferences: Our analysis is based on a simulated conjoint survey, which measures customers' stated preferences. While powerful, there can be a gap between what customers say they will do and their actual purchase behavior. A critical area for future work would be to validate and calibrate the WTP estimates from this model against real-world sales data, potentially creating a hybrid model that fuses survey insights with observed market outcomes to create a more accurate predictive tool.
2. Assumption of Independent Choices: Our current model follows the standard convention of assuming each choice a respondent makes is independent of their previous choices. In reality, factors like respondent fatigue or learning effects can occur during a survey. A more advanced model could incorporate time-series or state-space components to capture these potential dynamic effects within the survey-taking process.
3. Scope of Feature Interactions: The linear utility model (Equation 1) assumes that the value of each feature is additive. It does not account for potential feature interactions. For example, the perceived value of a "Pro" camera might be even higher when paired with 512GB of storage. Future work could explore more complex utility models that explicitly include interaction terms to capture these synergies.
4. Market Segmentation: Our hierarchical model captures individual-level preferences but does not explicitly segment the market into distinct personas (e.g., "Power Users," "Budget-Conscious"). Applying techniques like Latent Class Analysis or Dirichlet Process Mixture Models on top of our WTP estimates could automatically discover these customer segments, allowing for even more targeted product and marketing strategies [4].
5. Simplified Market Context: The simulation assumes a static competitive landscape. A significant area for future research would be to integrate competitor actions into the choice model. For instance, how does the WTP for an iPhone feature change when a new, compelling alternative from a competitor is introduced to the market?

Addressing these areas represents the next frontier in developing a truly comprehensive and dynamic system for product and pricing strategy. By tackling these challenges, we can build upon the foundation of valuation and uncertainty quantification established in this paper to create even more intelligent and responsive business decision-making tools.

## VII. CONCLUSION

The strategic pricing of product features is one of the most complex and high-stakes challenges faced by modern technology firms. This paper addressed this challenge by moving beyond simple heuristics and developing a rigorous statistical framework to answer the question: "What is a feature worth?"

We have demonstrated that a Bayesian Hierarchical Conjoint Model provides a powerful solution. By simulating a realistic consumer choice survey for a new iPhone, we successfully implemented a model that translates noisy, qualitative choices into precise, quantitative insights. The core contribution of our work is the direct estimation of Willingness-to-Pay (WTP) in clear, monetary terms, not as a single number, but as a full probability distribution that captures our uncertainty.

Our results were unequivocal: the model accurately recovered the true, underlying dollar value of key features like an upgraded camera system and a premium titanium frame. Furthermore, by extending the analysis to a revenue optimization simulation, we showed how these WTP distributions can be used to make a direct, data-driven pricing decision that maximizes expected revenue.

This research provides more than just a theoretical exercise; it offers a practical, end-to-end blueprint for any organization seeking to ground its product and pricing strategy in a scientific, data-driven foundation. By statistically deconstructing a product's price into the value of its parts, businesses can move from intuition to insight, making more intelligent decisions about what to build, how to price it, and ultimately, how to deliver maximum value to both their customers and their bottom line. Ultimately, this work demonstrates that the most valuable feature of any product is a price tag justified by data.